\documentclass[11pt,a4]{article}
\usepackage{amssymb}
\usepackage{amsmath}
\usepackage{amscd}
\usepackage{graphics}
\usepackage{graphicx}
\usepackage{exscale}

\begin{document}
\setlength{\textheight}{9.4in}
\setlength{\topmargin}{-0.6in}
\setlength{\oddsidemargin}{0.15in}
\setlength{\evensidemargin}{0.4in}
\renewcommand{\thefootnote}{\fnsymbol{footnote}}

\begin{center}
{\Large \bf The Structure of $D2$--Branes in the Presence of an RR field}

\vspace{0.4 cm}

{D.K. Park\raisebox{0.8ex}{1,2}\footnote[1]
{Email:dkpark@hep.kyungnam.ac.kr},
S. Tamaryan\raisebox{0.8ex}{1,3}\footnote[2]{Email:sayat@physik.uni-kl.de,
sayat@moon.yerphi.am}
 and
H.J.W. M\"uller--Kirsten\raisebox{0.8ex}{1}\footnote[4]
{Email:mueller1@physik.uni-kl.de}}

{\it 1. Department of Physics,
 University of Kaiserslautern, D-67653 Kaiserslautern, Germany\\
2. Department of Physics, Kyungnam University, Masan, 631-701, Korea\\
3. Theory Department, Yerevan Physics Institute, Yerevan--36,
375036, Armenia}
\end{center}

\vspace{1.8cm}

{\centerline{\bf Abstract}}
A Born--Infeld theory describing a $D2$--brane coupled to a
3--form RR potential is reconsidered and a
new type of static solution is obtained which is even stable.

\vspace{0.2cm}

\centerline{PACS Numbers: 11.10.Lm, 11.27.+d, 05.70.Fh}


\vspace{0.5cm}

\section{Introduction}
Recently in ref.\cite{1} the question was considered whether
a string can tunnel to a $D2$--brane in the presence of
a uniform background RR field, and it was shown that the string
can indeed nucleate the
spheroidal bulge of a $D2$--brane and can tunnel
to a toroidal $D2$--brane.  The tunneling was described by
bounces in Euclidean time and the action of these
entering the decay rate
of the string into a
toroidal $D2$--brane was deduced.  The transition process
was investigted in more detail in ref.\cite{2},
and the order of the quantum--classical transitions was
determined depending
on the magnitude of the
applied RR field.  All  D2-branes and the  
 Euclidean time tunneling configurations (i.e. bounces)
 considered
in these cases are unstable
as stated in ref.\cite{1},  and 
the latter therefore represent saddle points.
Here we are not concerned with the final states
of such tunneling, but rather with the
probability of tunneling away only as in ref.\cite{3}.

One can argue that depending on the potential,
a Euclidean time  tunneling 
 solution which is stable under small fluctuations
in its neighbourhood 
can  also exist as a  configuration with
 finite  action  and hence as a local  minimum of the
action  functional \cite{4}. 
 Our intention here is to search for 
such a  configuration  with finite energy.
 Since the
 transition rate exponent is proportional to 
minus the  value of the configuration action (below 
we use instead the
static energy), the tunneling
via Euclidean time  branes with higher
action is exponentially suppressed. Depending on the
action of such other configuration, the
qualitative tunneling consideration of ref.\cite{1} then
could or could not   be made
more  quantitative within the approximations
of the model.

The stability of Born--Infeld particles
has been studied in detail in refs.\cite{5,6} in the case
of the $D3$--brane (without an applied RR field).  In particular
it was shown there that the combined brane--antibrane
configuration is unstable, whereas the Born--Infeld string
is stable. The latter was also shown earlier in ref.\cite{7},
this being a consequence of the preservation of supersymmetry.

In ref.\cite{1} it was shown that an
initial state  Born--Infeld string in the
background of a uniform RR field
is unstable and can tunnel via Euclidean time
 saddle point brane configurations to some  $D2$--brane.
In the present work our original
intention was to search for   stable Euclidean time
 $D2$--branes with minimal action which would
therefore  dominate  other tunnelings.
Instead we found a configuration of higher
action than that of ref.\cite{1},  which,
however, is stable under small fluctuations
around it. This unusual property is
worth observing.

In Section 2 we
consider other brane
 configurations and show
that these are physically acceptable, i.e.
are nonsingular and have finite energy
(here we use the static solution in Minkowski
time instead of the equivalent Euclidean time configuration).
We
also show that for small values of the RR field there are two
types of solutions, one
type with lower  action (static energy) (used in \cite{1})
and a second, but stable  type with higher action.
In Section 3 we consider special solutions.
The equation for $D2$--brane configurations  admits 3 types of solutions:
Periodic solutions in terms of elliptic
functions, two constant solutions of cylindrical
shape, and solutions which are either
finite or vanish exponentially at infinity.
We are looking for branes with finite action, otherwise
the tunneling rate would be zero.
In particular we are considering the nucleation of
the unwrapped string so that
for $z\epsilon{\mathbf{R}}^1$ space only the third type
of solutions is physically acceptable.  Periodic
solutions can be used in the case of a
compactified space, when the wrapped
string tunnels into the toroidal
$D2$--brane \cite{2}.
In Section 4 we investigate the stability of such Euclidean
time tunneling branes
by  considering  the
fluctuation operator
 describing small deviations of the action in their  vicinity, and
demonstrate that in the case of our new
configuration this  has no negative
eigenvalue.  Vice versa it may possess a negative
eigenvalue for the other type of  solution and imply
that such a  configuration  is a saddle point.  In
Section 5 we conclude with some remarks.

\section{Formulation of the problem and conditions  for  static solutions
with finite  energy}

With the convention of
$\alpha^{\prime}=1$ the action of a $D2$--brane coupled with the
3--form gauge potential $A$ in Born--Infeld approximation
is given by \cite{1,8}
\begin{equation}
 I = -\frac{1}{4{\pi}^2g}
\int d^3\xi
\bigg\{\sqrt{-det\bigg(g^{ind}_{\alpha\beta} +
2\pi{\cal F}_{\alpha\beta}\bigg)}
+\frac{1}{3!}\epsilon^{\alpha\beta\gamma}A_{\mu\nu\rho}\partial_{\alpha}
X^{\mu}\partial_{\beta}X^{\nu}\partial_{\gamma}X^{\rho}\bigg\}
\label{1}
\end{equation}
where $\mu,\nu,\rho = 0,\cdot\cdot\cdot,9$ are spacetime indices, and
$\alpha,\beta,\gamma =0,1,2$ worldvolume indices and $g$ is
the string
coupling. The dilaton field is taken to be
constant
and the background field strength $H=dA$
is taken to be uniform and aligned with the brane, i.e.
$H_{0123}=h$=const. As in ref.\cite{1} we choose the world volume
to be cylindrical and hence define
\begin{equation}
X^0=t,\;\; X^1=z,\;\; X^2=R(t,z)\cos\sigma,\;\; X^3=R(t,z)\sin\sigma,
\;\; {\cal E}=2\pi F_{tz},
\label{2}
\end{equation}
and all other $X^i$ = const.
After integration over $\sigma$, the
action takes the form
\begin{equation}
I=\int dt \int dz {\cal L},\;\; {\cal L}= -\frac{1}{2\pi g}
\bigg(R\sqrt{1-{\dot R}^2-{\cal E}^2 + {R^{\prime}}^2} -\frac{h}{2}R^2\bigg)
\label{3}
\end{equation}
where dots and primes denote derivatives with respect to $t$ and $z$
respectively.
We observe that the 1-time-1-space combination $-{\dot R}^2+{R^{\prime}}^2$
is similar to that of 1+1 dimensional soliton theory, in which the
soliton of the static theory is identical with the instanton of
the Euclideanised 1-dimensional theory. In the following we
shall make use of this correspondence and consider instead of
the action  the energy of
the static configuration of the 2-dimensional theory. The time
dependence of the latter can be restored by Lorentz boosting. 
The canonical momentum
$D=2\pi g \delta I/\delta E$ must be
constant (cf. refs.\cite{1,2}) and
is given by $gn$, where $n$ is the number of
fundamental strings (cf. \cite{1,2}),
each of tension $1/2\pi$.
For static solutions the
energy $E$ is given by
\begin{equation}
E=\frac{1}{2\pi g}\int dz \bigg\{\sqrt{(1+{R^{\prime}}^2)(D^2+R^2)}
-\frac{h}{2}R^2\bigg\}
\label{4}
\end{equation}
Using the static expression (\ref{4}) instead of the proper
action (\ref{3}) for Euclidean time in the
computation of the decay rate, one has to subtract from the
action that of $n$ fundamental strings as emphasized in ref.\cite{1}.

Variation of $E$
with respect to $R$ yields
\begin{equation}
\frac{\delta E}{\delta R}-\frac{d}{dz}\frac{\delta E}{\delta R^{\prime}} =0
\label{5}
\end{equation}
which can be reduced to a first order
differential equation because it does
not contain the variable $z$ explicitly, resulting in
\begin{equation}
C = \sqrt{\frac{R^2+D^2}{1+{R^{\prime}}^2}}-\frac{h}{2}R^2, 
\label{6}
\end{equation}
where $C$ is a constant.  We rewrite this equation
\begin{equation}
R^{\prime}=\mp \frac{h}{hR^2+2C}\sqrt{(R^2_+-R^2)(R^2-R^2_-)},
\;\;R^{\prime}\neq 0
\label{7}
\end{equation}
with
\begin{equation}
{R_{\pm}}^2=\frac{2}{h^2}\bigg[(1-Ch)\pm\sqrt{1-2Ch+h^2D^2}\bigg]
\label{8}
\end{equation}
We observe that for physical reasons ($R$ has to be real)
$$
{R_-}^2\leq R^2\leq {R^2}_+
$$
which implies motion between these bounds, i.e. periodic motion.
Now we compare the expression (\ref{6}) for the integration
constant $C$ with expression (\ref{4})
for the energy density.  We observe that the
second (i.e. ``potential'') terms are the same
and the first (``kinetic'') terms differ by
a multiplicative factor, i.e. $(1+{R^{\prime}}^2)$,
and consequently
one may be led to believe that positive energy 
requires positive $C$.
Here we abandon this expectation  and consider
negative values of $C$. At first sight this may seem
dangerous since this suggests a pole in eq.(\ref{7})
but in spite of this
the solution is nonsingular
and the energy finite as will be shown below.
In this case eq.(\ref{7}) permits
solutions for all values of $h$ and for
$R^2_+$ and $R^2_-$ real and positive.  From eq.(\ref{8}) we
deduce two conditions
$$
C^2\geq D^2, \;\;\; 1-Ch\geq 0,
$$
which for positive values of $C$ imply
$$
hD\leq hC \leq 1
$$
meaning that $h$ must be less than the critical value
$h_C=1/D$.
On the contrary, for negative values of $C$, keeping in mind that
$h$ is positive, we have only the one condition that
$$
|C| \geq D
$$
(which excludes $R^2_{min}<0$)
so that $h$ can take any value, but $|C|$
is restricted to non--small values.  At the end of the next
Section we present a solution for large values of $h$, i.e.
when $ hD>>1 $.

We now demonstrate that solutions for
negative values of $C$ are, in principle, also acceptable,
i.e. are associated with a finite energy.
It is convenient in this case to set
\begin{equation}
C=-\frac{h}{2}a^2
\label{9}
\end{equation}
Then for $|C|\geq D$ we have
$$
0\leq R^2_-\leq a^2\leq R^2_+
$$
and $$
R^{\prime}=\mp \frac{\sqrt{(R^2_+-R^2)(R^2-R^2_-)}}{R^2-a^2}.
$$
We consider the approach $ R\rightarrow a$.  In this
domain$$
R^{\prime}\simeq \mp \frac{\lambda}{2}\frac{1}{R-a},\;\;
\lambda\simeq \frac{2\sqrt{a^2+D^2}}{ha}
$$
and, with integration constant $z_1$,
$$
(R-a)^2=\lambda|z-z_o|, \;\;
|R^{\prime}|=\frac{1}{2}\sqrt{\frac{\lambda}{|z-z_1|}}
$$
so that the crucial part of the energy (\ref{4})
around $z = z_1$  becomes
\begin{eqnarray}
& &\frac{1}{2\pi g}\int^{z_0+\delta}_{z_1-\delta}
dz\sqrt{(1+{R^{\prime}}^2)(D^2+R^2)}  \\  \nonumber
&\simeq& \frac{\sqrt{a^2+D^2}}{2\pi g}\int^{z_1+\delta}_{z_1-\delta}
|R^{\prime}|dz
=\frac{\sqrt{\lambda(a^2+D^2)}}{4\pi g}\int^{z_1+\delta}_{z_1-\delta}
\frac{dz}{|z-z_1|^{1/2}}
 < \infty
\label{10}
\end{eqnarray}
We conclude therefore that static solutions for negative values
of $C$ also have finite energy.

\section{Solutions with wheel-like shape}

We now consider some special solutions and their energy.
We consider first the special case $C=-D$ (i.e. $C<D$).
  In this case $D=ha^2/2,
R_-=0, R^2_+=4(1+hD)/h^2$.  Integrating the equation
$$
R^{\prime}=\mp\frac{R}{R^2-a^2}\sqrt{R^2_+-R^2}
$$
one obtains (note that $z=z_0$ corresponds to $R=R_+, z=+\infty$
to $R=0$ for the lower sign and $z=-\infty$
to $R=0$ for the upper sign)
\begin{equation}
\mp(z-z_0)=-\sqrt{R^2_+-R^2}+\frac{a^2}{R_+}\ln\frac{R_++\sqrt{R^2_+-R^2}}{R}
\label{11}
\end{equation}
for $z-z_0\lessgtr 0$ respectively.
In Fig. 1 we show the combination of these
two solutions, one representing the continuation (or
mirror image) of the other across the R-axis.
 One can clearly see that
the effect of negative values of $C$ is opposite
to that of positive $C$: Whereas positive $C$ yield an
elongated spheroidal bulge (cf. refs.\cite{1,2}),
 negative $C$ push this
bulging in the opposite direction
(thereby crossing the R-axis at a point
$R=R_0$ close to the origin)
 to eventually form a
wheel--like structure.
We  calculate the energy of this configuration
using eq.(\ref{6}) with $C$ there replaced by $-D$.
Substituting in the expression (\ref{4}) for the energy
the expression for  $(1+{R^{\prime}}^2)$ 
obtained from eq.(\ref{6}) one obtains
$$
E=\frac{1}{2\pi g}\int dz\bigg\{\frac{\pm(D^2+R^2)}
{C+\frac{h}{2}R^2}-\frac{h}{2}R^2\bigg\}.
$$
Here we have to choose the minus sign in order
to obtain a positive expression (even for the
divergent part, cf. the comment
after eq. (\ref{4})). With this choice and
rewriting the expression slightly, one has
$$
E=\frac{1}{2\pi g}\int dz\bigg\{
\frac{D(\frac{h}{2}R^2-D)-(R^2+\frac{h^2}{4}R^4)}
{\frac{h}{2}R^2-D}\bigg\}.
$$
This expression can now be rewritten as
$$
E= \frac{D}{2\pi g}\int dz
-\frac{1}{4\pi gh}\int^{\infty}_{-\infty}dz
\frac{R^2(4+h^2R^2)}{(R^2-a^2)}
= \frac{D}{2\pi g}\int dz +2E_-,
$$
where
$$
E_-=\frac{1}{4\pi gh}\int^{R_+}_{0} dR\frac{dz}{dR}
\frac{R^2(4+h^2R^2)}{(R^2-a^2)}.
$$
In this expression we now have to choose the
solution with the sign as in
$$
R^{\prime}=\frac{R\sqrt{R^2_+-R^2}}{R^2-a^2}
$$
(as in ref.\cite{1} this ensures that $R\rightarrow 0$
for $z\rightarrow \infty$).
With $D=+ng$ 
one now obtains 
\begin{equation}
E= \frac{n}{2\pi}\int dz +\frac{2R_+}{\pi gh}
+\frac{2h}{8\pi^2g}\bigg(\frac{4\pi}{3}R^3_+\bigg),
\label{12}
\end{equation}
whereas the energy of the spheroidal bulge of
ref.\cite{1} is
$$
E= \frac{n}{2\pi}\int dz 
+\frac{h}{8\pi^2g}\bigg(\frac{4\pi}{3}R^3_+\bigg).
$$
We observe that the energy of the bulge (the  term with $R^3_+$
which
results from $R^{\prime}\neq 0$) is here twice that of
the case considered in ref.\cite{1},
in addition to another term. Thus our
wheel-like solution has a higher energy.

Next we consider the case $C=-ha^2/2$ with $ ha>>1, hD>>1$. In this case
$$
R^2_\pm\simeq a^2\pm 2R_0, \;\; R_0\equiv \sqrt{a^2+D^2}/h, \;\; R_0<<a^2, D^2
$$
We set (with $\tau^{\prime}\equiv d\tau/dz$)
$$
R^2 \equiv a^2+\tau.
$$
Then eq.(\ref{7}) becomes
\begin{equation}
\frac{\tau^{\prime}}{2\sqrt{a^2+\tau}}=\mp\frac{\sqrt{4R^2_0-\tau^2}}{\tau}.
\label{13}
\end{equation}
Integration (for $\tau << a^2$) yields the equation ($R^2=x^2+y^2$)
\begin{equation}
z^2+\bigg(\frac{R^2-a^2}{2a}\bigg)^2=\frac{a^2+D^2}{h^2a^2}
\label{14}
\end{equation}
(with integration constant $z_0=0$).
This equation describes a deformed
circular structure with radius
$a$ (neglecting $2\sqrt{a^2+D^2}/h$) which is obvious if we look at its
intersection with the plane $z=0$.
 The structure has a thickness $\sqrt{a^2+D^2}/ha$.
Thus in the limit $ha>>1$ with
parameter $a$ fixed, the circular structure
becomes a shell of finite radius and small
thickness  as shown in Fig. 2.
A configuration like this  has been observed computationally in
ref.\cite{9} for a condition similar to the one here, i.e.
$C<0$..

\section{The stability of the solutions}

We now return to the question of whether the
Euclidean time tunneling  solutions considered
above are stable, i.e. are global minima of the action,
here considered as energy,
or not. For this reason we
consider the second variation $\delta^2E$ of the energy
in the vicinity of the classical solution.  Straightforward
calculation yields
\begin{equation}
\delta^2E =\frac{1}{4\pi g}\int\delta R{\hat M}\delta R dz
\label{15}
\end{equation}
where
\begin{equation}
{\hat M}=-\frac{d}{dz}Q\frac{d}{dz} + 2P\frac{d}{dz} +V
\label{16}
\end{equation}
with
\begin{equation}
Q=\frac{\sqrt{R^2+D^2}}{(1+{R^{\prime}}^2)^{3/2}}, \;\; P=\frac{RR^{\prime}}
{\sqrt{(1+{R^{\prime}}^2)(R^2+D^2)}}, \;\; V=D^2\frac{\sqrt{1+{R^{\prime}}^2}}
{(R^2+D^2)^{3/2}}-h
\label{17}
\end{equation}
A negative eigenvalue of the fluctuation
operator ${\hat M}$ implies
instability of the respective solution,  since the variation of
the solution in the direction of the corresponding
eigenfunction decreases the energy. Therefore it suffices to
investigate for which solutions
${\hat M}$ has only positive eigenvalues and for which
not.

One can derive another expression for
${\hat M}$ which is equivalent to the one above, i.e.
\begin{equation}
{\hat M} =
-\frac{1}{R^{\prime}}\frac{d}{dz}{R^{\prime}}^2Q\frac{d}{dz}
\frac{1}{R^{\prime}}
+2P\frac{d}{dz}-\frac{1}{R^{\prime}}(QR^{\prime\prime})^{\prime}
\label{18}
\end{equation}
but allows to present $\delta^2E$ as a sum
of positively defined terms plus a term
proportional to $C$.  On the basis of
this one can easily distinguish the stable solutions from
the unstable ones.

Here we are looking for branes with finite energy.
With $z\epsilon {\mathbf{R}}^1$
this implies the square integrability of the eigenfunctions
of ${\hat M}$.  The charge $D$ remains fixed by
quantisation (as stated earlier).
The positivity of all eigenvalues of ${\hat M}$
means that the mean value of ${\hat M}$ over any function
with finite norm is positive and vice versa.
We assume that $\delta R(z)$ is a
square integrable function, i.e.
\begin{equation}
\int \delta R(z)^2dz < +\infty,
\label{19}
\end{equation}
and consider the mean value of ${\hat M}$
on that class of functions.
This is the same as
 $\delta^2E$.
The function $R(z)$ is bounded, $R_-\leq R\leq R_+$,
 and $\delta R(z)$ must vanish at infinity.
Consequently we can integrate the term
$ 2\delta R P\frac{d}{dz}\delta R=P\frac{d}{dz}\delta R^2$ by parts and the
total derivative must vanish.  Also in the first term
we can return to the antihermitian operator
$d/dz$ to act on the term to the left yielding
a minus sign. As a result we have
\begin{equation}
\delta^2E=\frac{1}{4\pi g}\int
\bigg[{R^{\prime}}^2Q\bigg(\frac{d}{dz}\frac{\delta R}{R^{\prime}}\bigg)^2
+U\delta R^2\bigg]dz
\label{20}
\end{equation}
where
\begin{equation}
U=V-\frac{dP}{dz} -\frac{1}{R^{\prime}}\bigg(QR^{\prime\prime}\bigg)
\label{21}
\end{equation}

It is worth noting that one can
 make analogous manipulations with the differential
equation for eigenvalues of the operator ${\hat M}$.
After appropriate substitutions and eliminating the first order derivative, one
obtains the same expression (\ref{21}) for
the effective potential.  We prefer
the above way which allows us to connect
the second variation of the energy directly with the
integration constant $C$.
With some algebra one can deduce the explicit expression
for $U$:
\begin{equation}
U=h\frac{R^2}{R^2+D^2}+hD^2\frac{R^2}{(R^2+D^2)^2}-2C\frac{R^2}{(R^2+D^2)^2}
\label{22}
\end{equation}
The last term demonstrates in a transparent way the
stability of solutions with negative values
of $C$.  In contrast, in the case when $C>0$ and $hD^2\leq C$,
the negative
term becomes consequently a source for the instability which is
due to negative
eigenvalues.

\section{Conclusions}

In the above we obtained a new  nonsingular, finite--energy
(or rather action)  solution
of the Born--Infeld theory of a $D2$--brane in the presence of a
three--form RR--potential.
The wheel-like shape  of this solution appears
 as a  natural consequence of negative values
of the constant $C$ which imply  a threading of the
fundamental strings through its bulging in 
a direction which
is opposite to that of the spheroidal structure of ref.\cite{1}.
Also  naturally one expects the  more complicated
structure found here to have a higher energy.
  By considering small
fluctuations about a particular
wheel--shaped  configuration we also demonstrated its stability.

\vspace{4cm}

{\bf Acknowledgements:} D.K. P. and S.T. acknowledge support by the
Deutsche Forschungsgemeinschaft (DFG). S.T. also thanks S. Ketov
for discussions.


\begin{figure}[ht]
\centerline{\includegraphics[scale=1]{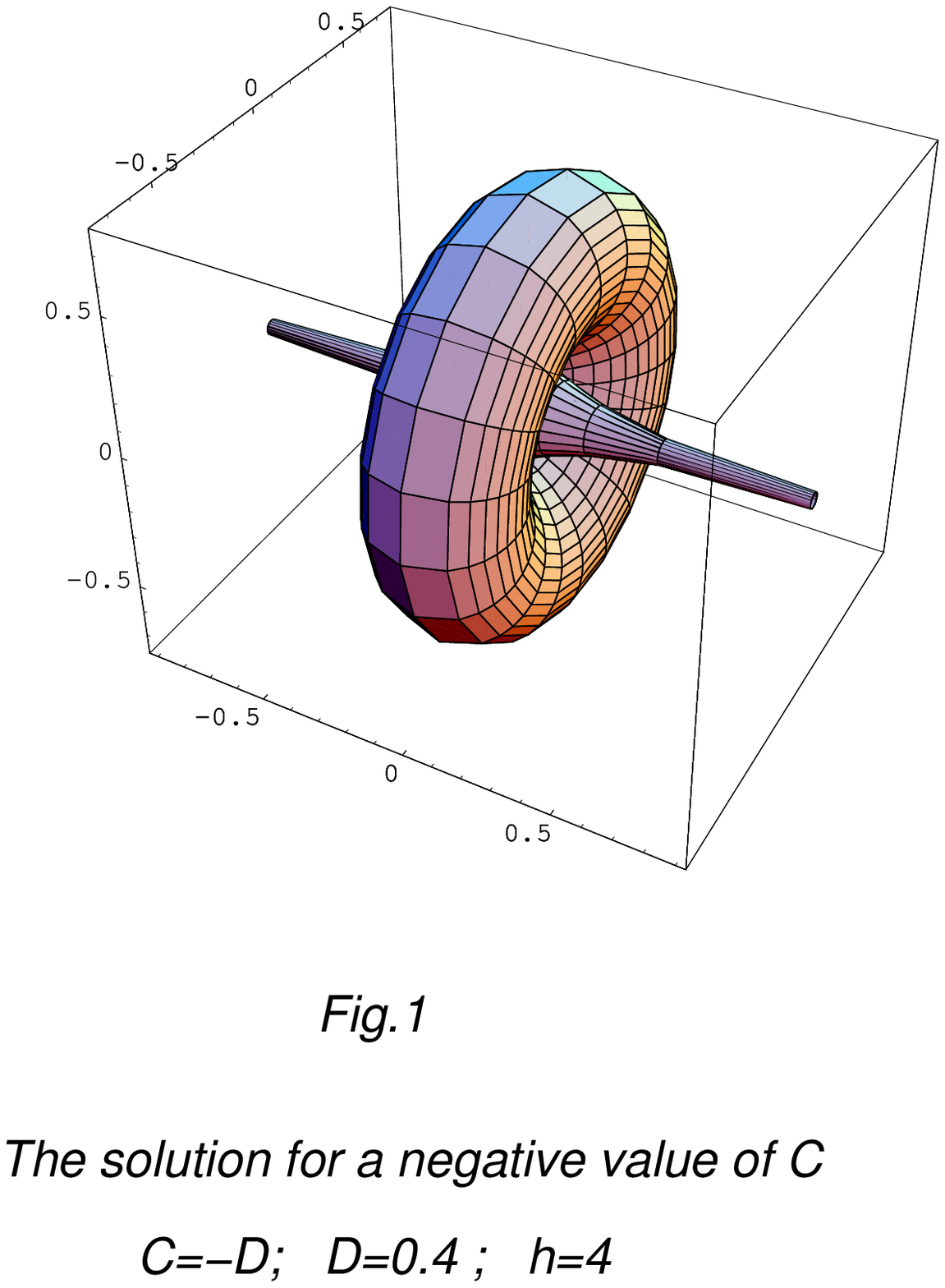}}
\end{figure}
\newpage
\begin{figure}[ht]
\centerline{\includegraphics[scale=1]{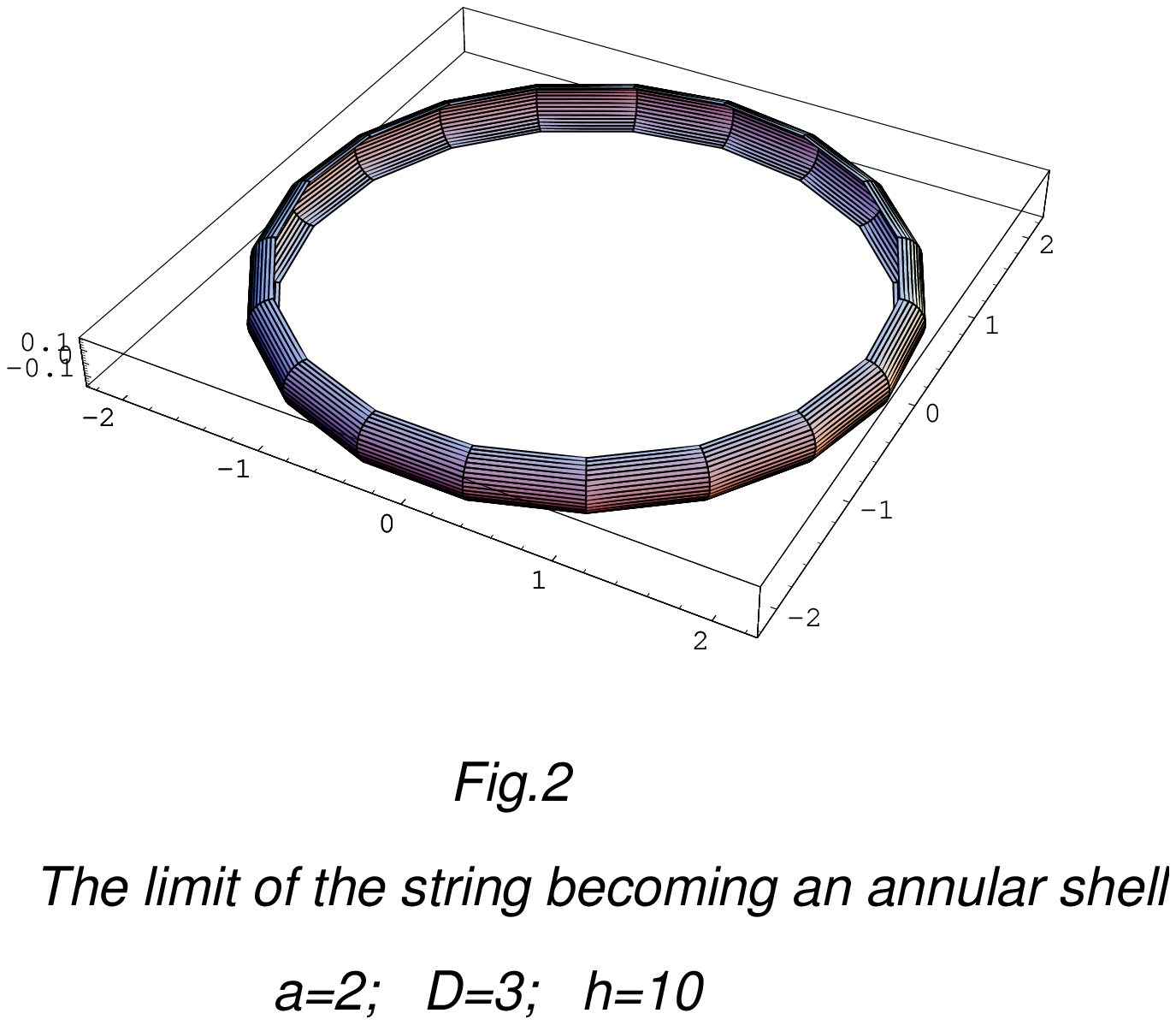}}
\end{figure}

\end{document}